\documentclass[aps,pre,reprint]{revtex4-1}

\usepackage{graphicx}
\usepackage{amssymb,latexsym,amsthm,amsmath,bm}
\usepackage{color}
\usepackage{ulem}

\begin{document}

\title{Emergent explosive synchronization in adaptive complex networks}

\author{Vanesa Avalos-Gayt\'{a}n}
\affiliation{Research Center in Applied Mathematics, Univ. Aut\'{o}noma de Coahuila, Saltillo, Coahuila, Mexico}

\author{Juan A. Almendral}
\affiliation{Center for Biomedical Technology, Univ. Polit\'ecnica de Madrid, 28223 Pozuelo de Alarc\'on, Madrid, Spain}
\affiliation{Complex Systems Group \& GISC, Univ. Rey Juan Carlos, 28933 M\'ostoles, Madrid, Spain}

\author{I. Leyva}
\affiliation{Center for Biomedical Technology, Univ. Polit\'ecnica de Madrid, 28223 Pozuelo de Alarc\'on, Madrid, Spain}
\affiliation{Complex Systems Group \& GISC, Univ. Rey Juan Carlos, 28933 M\'ostoles, Madrid, Spain}

\author{F. Battiston}
\affiliation{School of Mathematical Sciences, Queen Mary University of London, 
London E1 4NS, United Kingdom}

\author{V. Nicosia}
\affiliation{School of Mathematical Sciences, Queen Mary University of London, 
London E1 4NS, United Kingdom}

\author{V. Latora}
\affiliation{School of Mathematical Sciences, Queen Mary University of London, 
London E1 4NS, United Kingdom}

\author{S. Boccaletti}
\affiliation{CNR-Institute of Complex Systems, Via Madonna del Piano, 10, 50019 Sesto Fiorentino, Florence, Italy}
\affiliation{Embassy of Italy in Israel, Trade Tower, 25 Hamered St., 68125 Tel Aviv, Israel}

\begin{abstract}
Adaptation plays a fundamental role in shaping the structure of a complex network and improving its 
functional fitting. Even when increasing the level of synchronization in a biological system is considered as the main 
driving force for adaptation, there is evidence of negative effects induced by 
excessive synchronization. This indicates that coherence alone can not be enough to explain all the 
structural features observed in many real-world networks. 
In this work, we propose an adaptive network model where the dynamical evolution of the node states 
towards synchronization is coupled with an evolution of the link weights based on an anti-Hebbian adaptive rule, which accounts 
for the presence of inhibitory effects in the system.
We found that the emergent networks
spontaneously develop the structural conditions to sustain explosive
synchronization. Our results can enlighten the shaping mechanisms at the heart
of the structural and dynamical organization of some relevant biological systems, namely brain networks, for which the emergence of
explosive synchronization has been observed.
\end{abstract}

\maketitle

\section{Introduction}

Adaptivity is a key feature in the construction and function of many
real complex systems. In the brain, for instance plasticity is at the
heart of memory and learning processes, and governs the huge
functional versatility of this system. In many modeling approaches,
synchronization has been studied as the emerging collective
phenomenon of interest in a population of interacting dynamical units,
and the target of adaptation has been considered to be the improvement
of the level of synchronization in the system \cite{Zhou2006,DeLellis2008}.
Most of the works in this area have implemented different versions of
the Hebbian adaptation rule as a generative mechanism to enhance the strengths
of the links of a network \cite{Assenza2011,Gutierrez2011, Avalos2012}.
As a result, the origin of various emerging features of real-world networks, 
such as scale-free topologies \cite{Yuan2013}, modularity
\cite{Assenza2011,Gutierrez2011}, or assortativity \citep{Avalos2012}, 
has been better understood.

However, even if brain synchronization, both at the scale of neurons
and at that of cortical areas, has proved to be fundamental for the
proper functioning of the brain, improving the coherence cannot be the
only mechanism at work in brain networks. In fact, it is well known
that oversynchronization can destroy the overall complexity of the
system, reducing the amount of information that the system is able to
process and eventually leading to pathological states as epilepsy 
\cite{Chavez2010}. For this reason, inhibition and anti-Hebbian
coupling have been investigated in neural systems, and they have been
shown to play an important role in the control of excessive
synchronization and redundancy \cite{Lamsa2007, Harvey2010,
  Feldman2012}, and also in the context of circadian rhythms
\cite{Myung2015}. Anti-Hebbian rules have also been considered more in
general in adaptive complex networks, and it has been found that they
can be useful to generate features as criticality \cite{Magnasco2009},
dissasortativity \cite{Yuan2013}, structural heterogeneity
\cite{Scafuti2015}, bistability \cite{Skardal2014} or multistability
\cite{Chandrasekar2014}.

In this work, we introduce and study an adaptive complex network model in which
the nodes are coupled oscillators trying to synchronize their phases,
while the link weights evolve with an anti-Hebbian rule which weakens
the connection between pairs of coherent nodes. 
We will show that the competition
between the attractive coupling and the anti-Hebbian link dynamics generates 
networks with interesting topological and dynamical features.
In particular, we find that the networks produced by our model are able to
sustain explosive synchronization (ES), 
and this happens for large range of the two tuning parameters of the model. 
Explosive synchronization, i.e. the sudden, discontinuous and
irreversible transition from an incoherent state to a fully
synchronized one, is a phenomenon that has attracted special attention
in last few years
\cite{Gardenes2011,Leyva2013a,Leyva2013b,Navas2015,explosive_report,Zhang2015}.
Explosive synchronization has been also observed experimentally in
 power grids \cite{Motter2013}, circuits \cite{Leyva2012} and chemical
 reactions \cite{Kumar2015}. More recently, the interest has extended
also to neuroscience, and some experimental studies have related
explosive synchronization to the onset of seizures \cite{Yaffe2015}, and the
transition to and from consciousness in anaesthetized patients
\cite{Kim2016,Kim2017}. 
Our results indicate that the frustration induced
by an anti-Hebbian adaptation rule that  
promotes links between nodes in anti-phase dynamics,
can be the main responsible of the phenomena observed empirically
in neuroscience.
Thus, having a simple model that produces networks able to suddenly
switch to full coherence, can turn very useful to explore the role
of the different tuning parameters and to understand the basic
mechanisms behind explosive synchronization in neural systems.

The article is organized in the following way.
In Section II we introduce our new coevolving network model coupling node synchronisation and anti-Hebbian pairwise interactions. 
In Section III we show by mean of numerical simulations the basic structural and dynamical feature of the emerging networks, and how they sustain explosive synchronization.
In Section IV we consider the simplified version of the model with only
two coupled oscillators, which is amenable to analytical solution and
allows to grasp the basic mechanisms at work in the considered coevolving process also for larger coupled systems.
Finally we draw our conclusions in Section V.

\section{The model}

We consider a system of $N=300$ all-to-all coupled Kuramoto
oscillators, the simplest and most common way to describe
synchrony in nature \cite{Kuramoto1984,Boccaletti2002}.
Each oscillator is characterized by its phase $\theta_l$, with $l=1,\dots, N$, 
whose dynamics is ruled by the equations:
\begin{equation}
    \dot{\theta}_l = \omega_l + \frac{\sigma_c}{N} \sum_{m=1}^N
    \alpha_{lm} \sin(\theta_m - \theta_l), \label{fun:inst-phases}
\end{equation}
where $\omega_l$ is the natural frequency associated to the oscillator, that 
is assigned randomly from a uniform distribution in the interval $[0.8, \,
  1.2]$, $\alpha_{lm} \in [0, \, 1]$ is the weight of the connection
between the units $l$ and $m$, and $\sigma_c$ is the so-called {\it coupling
  strength}, the control parameter that allows to tune the strength of the
interactions. 
Despite the original model was not devised to describe neurons or
groups of neurons, in its simplicity, the model is able to capture the
gist of a synchronous behavior, which explains why it is nevertheless
useful for the investigation of brain networks
\cite{Nicosia2013,Maksimenko2017}.
In our work we consider the case in which the weight of a link 
can differ from one pair of nodes to another. Furthermore, we
assume that the weights of the links can change in time according
to the dynamics of the corresponding two end nodes. 
Namely, the quantities $\alpha_{lm}$, with $l,m=1,\dots, N$,
are considered to be time dependent variables, $\alpha_{lm} = \alpha_{lm}(t)$, 
obeying a differential logistic equation, with a growth rate which depends on
the correlations of the two corresponding oscillators $l$ and $m$. 
The equations read: 
\begin{equation}
    \dot{\alpha}_{lm} = (p_c - p_{lm}) \ \alpha_{lm} (1 - \alpha_{lm}), \label{fun:weight}
\end{equation}
where $p_{lm}=p_{lm}(t)$ is the instantaneous phase correlation between units $l$ and
$m$ at time $t$, defined as: 
\begin{equation}
    p_{lm}(t) = \frac{1}{2} \left| e^{i \theta_l(t)} + e^{i
      \theta_m(t)} \right|,\label{fun:correlation}
\end{equation}
and $p_c$ is a {\it correlation threshold}. The threshold is the
second tuning parameter of our model, and has the following
meaning. Whenever $p_{lm}(t) < p_c$, the weight of the link $(l,m)$
gets increased by the dynamical evolution of Eq.~(\ref{fun:weight}),
while the weight $\alpha_{lm}$ is decreased when $p_{lm}(t) > p_c$.
Thus, once the value of $p_c$ is fixed, the links connecting pairs 
of oscillators with a higher (lower) level of instantaneous synchronization
will be weakened (reinforced). As we will show, 
such a mechanism of frustration of the local synchronization
process, which affects a larger number of pairs the higher is the
value of $p_c$, is an essential ingredient for the emergence of 
explosive synchronization. 
Notice that Eq.~(\ref{fun:weight}) is bistable, and therefore each
weight $\alpha_{lm}(t)$ will tend to converge in time to either one of the two
values, 0 or 1.  As a consequence, for any given choice of the two
tuning parameters of our model, $\sigma_c$ and $p_c$, and a random
sampling of the initial conditions ${\theta}_l(t=0)={\theta}_l(0)$, 
and $\alpha_{lm}(t=0)=\alpha_{lm}(0)$, the coevolution of oscillator states
and link weights governed by Eq.~(\ref{fun:inst-phases}) and Eq.~(\ref{fun:weight})
will result in a progressive pruning of the links of the
initally complete graph, until the system reaches an asymptotic state, defined by $\dot{\alpha}_{lm}=0$, $\forall \, l, m$,
corresponding to a specific dynamically-induced network
topology.

The phase correlation introduced in Eq.~(\ref{fun:correlation}) is an
instantaneous measure of the correlations between two nodes, and it does not 
depend on any long-term synchronization
process \cite{Avalos2012}. This represents a qualitative difference
between the present study and that conducted in
Refs.~\cite{Assenza2011,Gutierrez2011}, where it was shown that memory
dependent adaptation mechanisms may induce 
the simultaneous appearance of synchronized clusters and 
scale-free distributions for the
weights of a network. Here, instead, is the adaptive nature of the
interactions that directly shapes the topology of a network, and all what
will be discussed hereinafter will be about the connectivity 
properties of the emergent network structure.

\section{Results}

For a large range of parameters $\sigma_c$ and $p_c$ we measure the final degree of global synchronization by means of the Kuramoto order parameter
\[ \textstyle R := \left\langle \frac{1}{N} \left| \sum_{l=1}^N e^{i \theta_l(t)} \right| \right\rangle_t, \]
with $R \approx 1$ indicating a fully synchronized graph, and $R \approx 0$ an asynchronous behavior. All the measures are averaged over 10 different instances of the system.

\begin{figure}[htb]
    \includegraphics[width=0.23\textwidth]{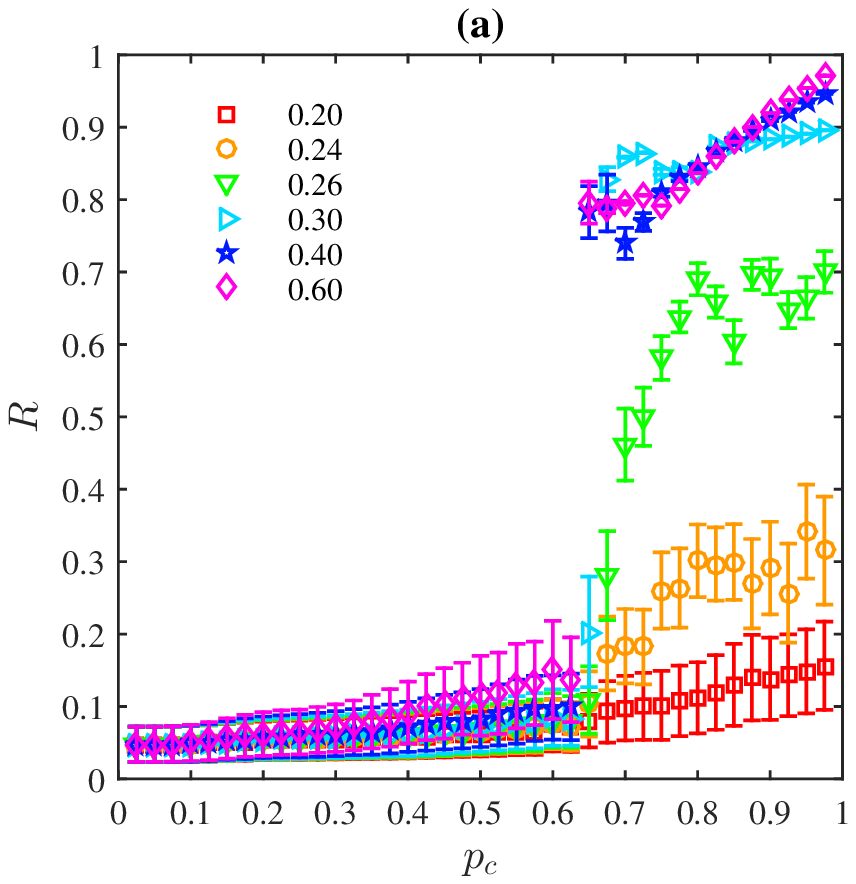}
    \includegraphics[width=0.23\textwidth]{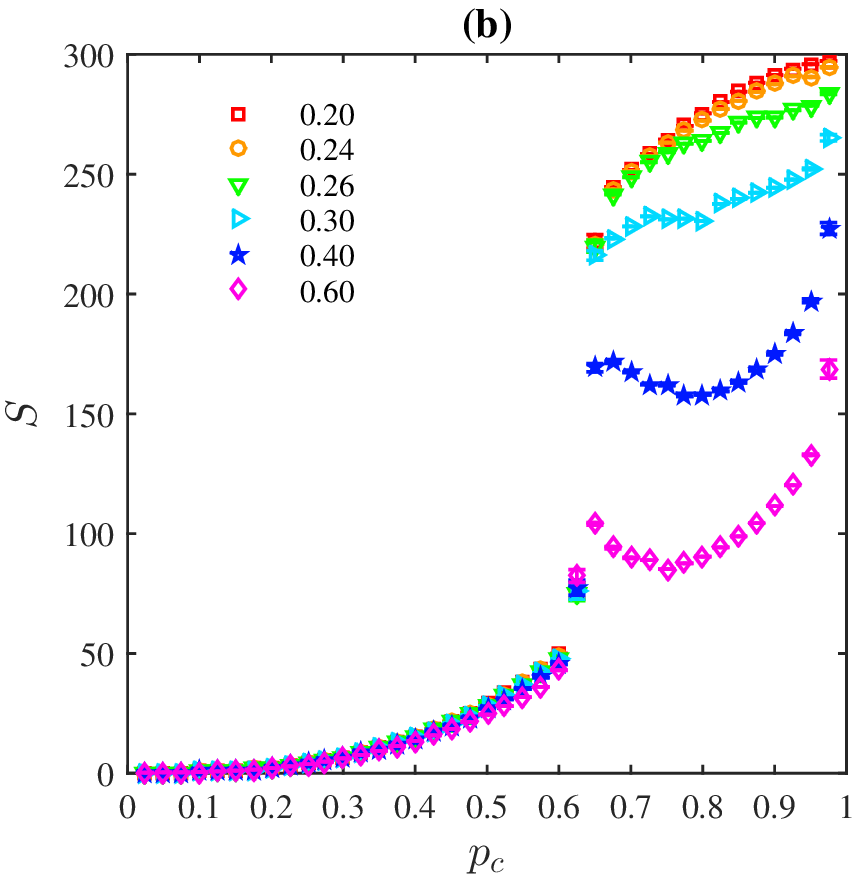}
    \caption{Asymptotic values of (a) global synchronisation, $R$, and (b) 
      total strength, $S$, in coevolving weighted networks governed
      by Eqs.~(\ref{fun:inst-phases}) and (\ref{fun:weight}),  
      as a function of the correlation threshold 
      $p_c$, and for different values of the coupling strength $\sigma_c$. 
      When $\sigma_c>\sigma_c^* \simeq 0.255$,
      abrupt transitions indicating the emergence of explosive
      synchronization are observed at $p_c \simeq 0.625$.}
    \label{fig:global_weig}
\end{figure}

\begin{figure}[htb]
    \centering \includegraphics[width=0.45\textwidth]{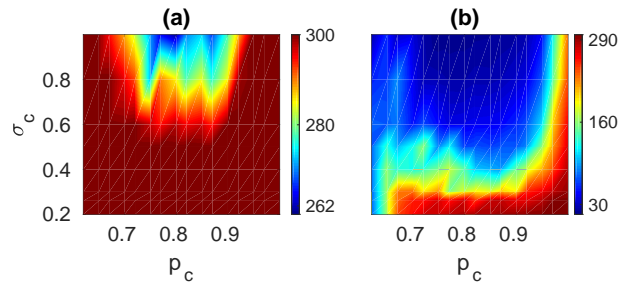}
    \caption{(a) Size of the giant connected component, $N_G$, in the
      asymptotic binarized networks and (b) average
      degree $\langle k \rangle$, as a function of the coupling
      strength $\sigma_c$ and of the correlation threshold $p_c$.}
    \label{fig:structural}
\end{figure}

In Fig.~\ref{fig:global_weig}(a) we characterize the asymptotic
dynamics of the network by reporting the average value of $R$ as a
function of $p_c$ for different values of $\sigma_c$. As expected, for
small values of the coupling $\sigma_c$ the network is not able to
synchronize for any value of $p_c$. However, above the critical
coupling for the Kuramoto model $\sigma_c^* =\frac{2}{\pi
  g(0)}=0.8/\pi \simeq 0.255$ \cite{Acebron2005}, the global
synchronization shows an abrupt transition at $p_c \simeq 0.625$. For
slightly smaller values of $\sigma_c$, the global strength suffers
also a sharp transition, as shown in Fig.~\ref{fig:global_weig}(b),
pointing out to the fact that the networks go through a phase of
strong local synchronization before global coherence is achieved.
To better characterize the properties of the emerging networks, in the
following we analyse the adjacency matrices $A=\lbrace a_{lm} \rbrace$
associated to the original weighted graphs $W= \lbrace \alpha_{lm}
\rbrace$, where we set $a_{lm}=1$ if $\alpha_{lm} > \tau$, and
$a_{lm}=0$ otherwise. We choose $\tau=0.8$ as a very conservative
threshold to ensure that we keep all the significant links.  The
system is explored for $p_c = [0.6, 1]$ and $\sigma_c =[0.2, 1]$, the
relevant parameter range deduced from Fig.~\ref{fig:global_weig}.
As the thresholding process can lead to node pruning, we have
performed a component analysis and computed the size $N_G$ 
of the largest connected component. Results are again averaged over
10 different realization of the process. 
Fig.~\ref{fig:structural}(a) shows a heat map of
the average size of the giant connected component
$N_G(\sigma_c,p_c)$ for each pair of 
control parameters $(\sigma_c, p_c)$. We notice that
in a large area of the phase diagram,
all the $N=300$ nodes belong to a single component
However, when measuring the average degree
$\langle k \rangle$ in Fig.~\ref{fig:structural}(b), we observe that
the network connectivity decreases from that of fully connected networks
to $\langle k \rangle =2$ for large values of $\sigma_c$.

\begin{figure}[htb]
    \begin{center}
    \includegraphics[width=0.52\textwidth]{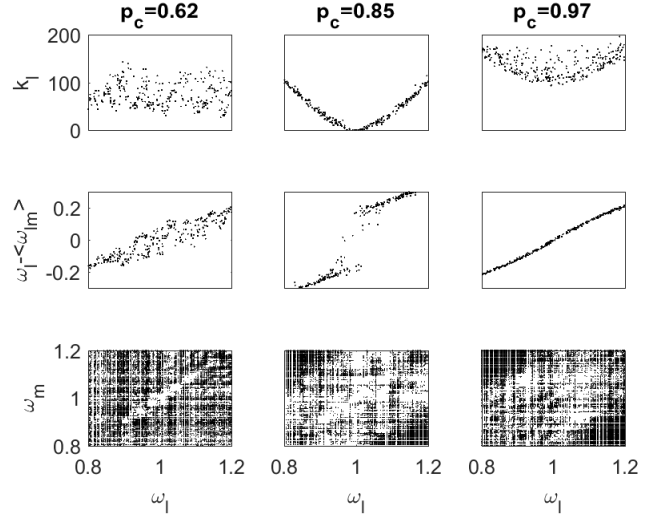}
    \end{center}
    \caption{Scatter plot of the node degree $k_i$ (upper row panels)
      and of the node neighborhood detuning $\omega_i - \langle
      \omega_{im} \rangle$ (central row panels) as a function of the
      natural frequency $\omega_i$ of a node for $\sigma_c=0.6$. The connectivity of the resulting networks are shown as matrices
      where nodes are placed according to their natural frequencies $\omega$ (bottom row
      panels).}
    \label{fig:micro}
\end{figure}

A more interesting information is retrieved when the correlations between the microscopic features are inspected. In Fig.~\ref{fig:micro} we show for three representative values of $p_c$, namely $0.62$ (left), $0.87$ (center) and $0.97$ (right panels), for a fixed value $\sigma_c =0.6$ -high enough to enable synchronization as shows Fig.~\ref{fig:global_weig}(a)-, how the emergent network structure and its dynamics are intertwined. In the first row of Fig.~\ref{fig:micro} the degree of each node $k_l$ is pictured as a function of the frequency $\omega_l$. Before the transition ($p_c=0.62$), the two features are uncorrelated. However, after the transition, a strong correlation between $k_l$ and $\omega_l$ appears, meaning that the nodes with $\omega_l$ at the two edges of the natural frequency distribution are much more connected. This is a surprising and noticeable feature, as  V-shape k-$\omega$ relationships are well known to be characteristic of networks capable of sustainaing explosive synchronization \cite{Leyva2013a,Leyva2013b,Navas2015,explosive_report}. The finding is confirmed by the fact that the dynamics in Eq.~(\ref{fun:weight}) forces the network to acquire {\it frequency dissasortativity}, i.e., nodes are much more likely to link to those with distant frequencies, as can be seen in the central row of Fig.~\ref{fig:micro}, where we plot the frequency detuning of each node with its averaged neighbor frequency $\omega_l-\langle \omega_m \rangle_l$, where $m \in  {\cal G}_{l}$ are the nodes in the neighborhood of node $l$. The phenomenon is especially striking just after the transition (center panel). In addition, we also see that for intermediate values of $p_c$, the detuning is bistable inside the same network.

These results show that the link evolution has actually reinforced links connecting nodes whose frequencies are as far as possible, progressively pruning the remaining connections. This process can be followed through the panels in the bottom row of Fig.~\ref{fig:micro}, where the full connectivity matrix is plotted as a function of ($\omega_l,\omega_m$). Before the transition to synchrony (first column, $p_c$=0.62), the pruning affects only nodes with frequencies close to the center of the distribution, as can be seen in the left panel. As $p_c$ grows, the link suppression affects a larger number of links connecting nodes with higher detuning, and therefore $\langle k \rangle$ and eventually $N_g(\sigma_c,p_c)$ decrease. A further increase of $p_c$ generates larger values of the mean strength in the network, Fig.~\ref{fig:global_weig}(b), and a slight increase in the number of links above the threshold, as shown in Fig.~\ref{fig:micro} (right bottom panel).

The structural inspection of the networks obtained through the adaptive process described by Eqs.~(\ref{fun:inst-phases}) and (\ref{fun:weight}) has revealed that for a wide range of the tuning parameters the emerging systems present a strong frequency-degree correlation and frequency dissasortativity. This hints that such networks could be able to sustain first-order synchronization, here critically controlled by the correlation threshold $p_c$.

We numerically check this prediction by using the resulting networks as the fixed connectivity support of a system of interacting Kuramoto oscillators, described by the following dynamical equation
\begin{equation}
    \dot{\theta}_l = \omega_l + \frac{\sigma}{N(\sigma_c,p_c)} \sum_{m=1}^{N(\sigma_c,p_c)} a_{lm} \sin(\theta_m - \theta_l), \label{fun:new_kura}
\end{equation}
where $l=1, \dots, N(\sigma_c,p_c)$, being $N(\sigma_c,p_c) \leq N$ the size of the binarized network obtained after the adaptive process for parameters $(\sigma_c,p_c)$, and $A' =\lbrace a_{lm} \rbrace$ the adjacency matrix of the corresponding largest giant component of size $N_g$. The coupling strength $\sigma$ is now set to be the only control parameter, regardless of the original value of $\sigma_c$ used to create $A'$. The original frequency $\omega_l$ of each node $l$ is maintained to preserve the structure-dynamics correlations which appeared as a result of the adaptive process. The coherence $R$ is monitored as a function of $\sigma$ by gradually increasing its value along the simulation without resetting the system. As long as we are looking for a possible first-order phase transition, and the expected corresponding synchronization hysteresis, we perform the simulations also in the reverse way: i.e. we start from a given value $\sigma_{max}$ where the system is found to be synchronized, and gradually decreasing the coupling. In the following, the two sets of numerical trials are termed as {\it forward} and {\it backward} branches, respectively, as used in the literature \cite{Gardenes2011,explosive_report}.

\begin{figure}[htb]
    \centering
    \includegraphics[width=0.45\textwidth]{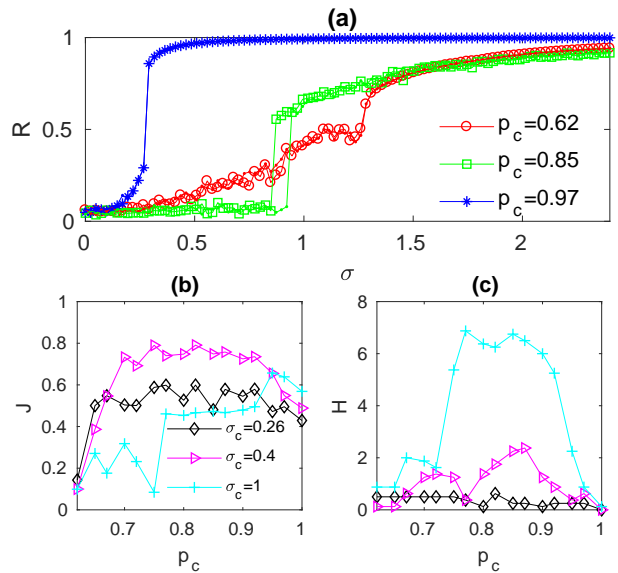}
    \caption{(a) Global synchronisation $R$ for forward and backward synchronization schemes for the binary networks considered in Fig.~\ref{fig:micro}. In panel (b-c) we investigate the abrupt nature of the synchronization transition by reporting the maximum synchronization jump $J$ (b) and the width of the hysteresis loop $H$ (c) as a function of the coupling strength $\sigma_c$ and of the correlation threshold $p_c$.}
    \label{fig:dynamical}
\end{figure}

In Fig.~\ref{fig:dynamical}(a) we show the synchronization schemes for the same examples whose structures have been studied in Fig.~\ref{fig:micro} ($\sigma_c=0.6$; $p_c=0.62, 0.85, 0.97$). The three types of structures correspond indeed to different dynamical behaviors. For networks obtained with $p_c$ below the transition, where structural-dynamics correlations are not observed, the network synchronizes in a second order, reversible transition (red circles). For $p_c$ values  beyond the critical value, the coherence is reached through a first order transition, as suggested by the strong frequency-degree correlations. For intermediate values (green squares), this transition to synchrony is very delayed, and presents an hysteresis loop as a consequence of the bi-stability generated in the detuning distribution (already commented in Fig.~\ref{fig:micro}, central panel). For even  larger $p_c$ values (blue diamonds), corresponding to networks with softer k-$\omega$ correlation and without gap bi-stability, the synchronization occurs abruptly but reversibly.

A wider overview of how the tuning parameters $\sigma_c$ and $p_c)$ determine these dynamical behaviors is shown in Fig.~\ref{fig:dynamical}(b-c). As a measure of abruptness of the synchronization transition, in Fig.~\ref{fig:dynamical}(b) we compute the average value of the maximum difference in the value of $R$ for two consecutive values of the coupling parameter
\[ \textstyle J := \max_{\sigma} \{ R(\sigma + \delta \sigma)-R(\sigma) \} \]
for the forward transition. Similar results are found for the backward transition. In Fig.~\ref{fig:dynamical}(c), we measure the width of the hysteresis loop by computing the distance between the critical synchronization couplings for the backwards ($\sigma_b$) and the forward ($\sigma_f$) processes,
\[ \textstyle H := \sigma_{b}-\sigma_{f}. \]
As can be seen, the parameter region where the width of the hysteresis loop presents significant values closely corresponds to the values where $N_g(\sigma_c,p_c)<N$, Fig.~\ref{fig:structural}(a), when the frustration for the nodes with frequencies close to the center of the distribution is so strong that they eventually become disconnected and are removed from the binary network.

\section{Analytical analysis of a simple adaptive system}

To better understand the dynamical system considered in the previous sections, we proceed to analytically study of two oscillators $\theta_1$ and $\theta_2$ coupled by a single weighted link, $\alpha = \alpha_{12} = \alpha_{21}$. For this simplified system, Eqs.~(\ref{fun:inst-phases}) and (\ref{fun:weight}) reduce to
\begin{eqnarray*}
\dot{\theta}_1 &=& \omega_1 + \frac{\sigma_c}{2} \ \alpha \sin(\theta_2 - \theta_1), \\
\dot{\theta}_2 &=& \omega_2 + \frac{\sigma_c}{2} \ \alpha \sin(\theta_1 - \theta_2), \\
\dot{\alpha} &=& (p_c - p) \ \alpha (1 - \alpha),
\end{eqnarray*}
where $p = p_{12} = p_{21}$ is the instantaneous phase correlation between oscillators 1 and 2 defined in Eq.~(\ref{fun:correlation}),
\[ p = \sqrt{ \frac{1+ \cos (\theta_1 - \theta_2)}{2}. } \]

Without loss of generality, we suppose that $\Delta := \omega_1 - \omega_2 >0$ in order to transform the former set of equations into a two dimensional system with variables $\alpha$ and the phase difference between oscillators $\phi := \theta_1 - \theta_2$, so that we have
\begin{eqnarray}
\dot{\phi} &=& \Delta - \sigma_c \ \alpha \sin \phi,  \label{fun:2nodes} \\
\dot{\alpha} &=& \left( p_c - \sqrt{ \frac{1+ \cos \phi}{2} } \right) \alpha (1 - \alpha), \nonumber
\end{eqnarray}

As it is not possible to integrate Eqs.~(\ref{fun:2nodes}) explicitly, we analyse the system stability to grasp insights on its qualitative behavior. The details of this calculation can be found in the Appendix. We summarize all the results in Fig.~\ref{fig:regions}, which depicts the dynamical behaviors for a system described by Eqs.~(\ref{fun:2nodes}) as a function of the usual coupling strength $\sigma_c$ and the correlation threshold $p_c$. The color map of Fig.~\ref{fig:regions} represents the asymptotic value of the weighted link, as calculated in the Appendix.

Regions $A_{1,2}$ correspond to those parameters for which there is no phase locking: $A_1$ is defined by $p_c < 1/\sqrt{2}$, and $A_2$ by $p_c > 1/\sqrt{2}$ and $\sigma_c < \Delta$. The main difference between these two regions is that whereas in $A_1$ the weight tends to zero, so that the oscillators end up being disconnected, in $A_2$ the link weight tends to one, which is not enough to synchronize the oscillators since $\sigma_c < \Delta$.
\begin{figure}[htb]
    \centering
    \includegraphics[width=0.45\textwidth]{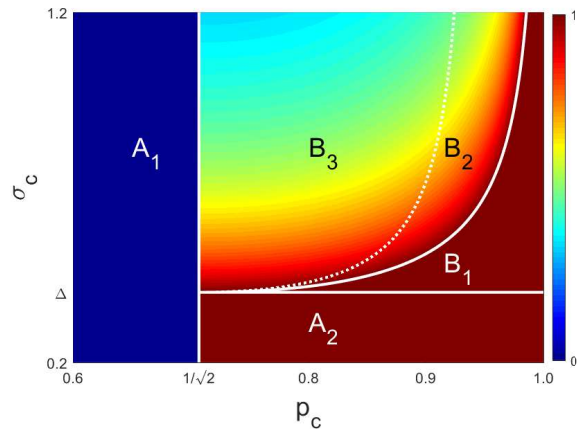}
    \caption{Stability diagram for the link weight of a system of two coupled oscillators following Eqs.~(\ref{fun:2nodes}). The link weight
    tends to $0$ in region $A_1$ and to $1$ in region $A_2$, but in both cases synchronization is impossible. Regions $B_{1,2,3}$ are characterised by phase locking: in $B_1$ the weight becomes one, whereas in $B_{2,3}$ the weight becomes $\Sigma_1/\sigma_c <1$ with $\Sigma_1$ defined in Eq.~(\ref{fun:sigma1}). Regions $B_2$ and $B_3$ are different as the latter has a spiral sink, resulting in a slower convergence to the asymptotic state.}
    \label{fig:regions}
\end{figure}

On the other hand, we have synchronization in regions $B_{1,2,3}$. $B_1$ is defined by  $\Delta<\sigma_c < \Sigma_1$, with the critical curve $\Sigma_1$ defined in Eq.~(\ref{fun:sigma1}). $B_2$ is enclosed by $\sigma_c > \Sigma_1$ and $\sigma_c < \Sigma_2$, with $\Sigma_2$ defined in Eq.~(\ref{fun:sigma2}). Finally, $B_3$ is characterized by $\sigma_c > \Sigma_2$ and $p_c > 1/\sqrt{2}$. Whereas in $B_1$ the weight becomes one, in $B_{2,3}$ tends to $\Sigma_1/\sigma_c <1$, where $\Sigma_1$ is calculated in Eq.~\ref{fun:sigma1}. The difference between $B_2$ and $B_3$ is that we have a sink node in the first region and a spiral sink in the second one, which implies that the convergence in $B_3$ is slower than in $B_2$.

Even for this simple two-node analysis, the scenario displayed by Fig.~\ref{fig:regions} fits closely the results obtained by numerical analysis for the evolution of large scale adaptive networks. This can be seen by comparison with Fig.~\ref{fig:structural}(b), where the size of the giant component, which directly depends on the stationary weight values, was shown. In particular, the presence of the same structure in the parameter space allows us to consider that the analysis captures the relevant microscopic dynamics at the heart of the entanglement between structure and dynamics observed in the network resulting from the coevolution process.

\section{Conclusions}

In conclusion, in this work we suggested a novel adaptive network model based on the competition
between attractive coupling at the node level and anti-Hebbian repulsive dynamics at the link level.
We showed how an initial set of fully interacting phase oscillators can naturally evolve towards a complex networked system, under the action of an adaptation mechanism which promotes interactions between elements that are not synchronized.  Indeed, in many biological systems synchronization needs to be both promoted and controlled in order to avoid excessive redundancy. 
We characterized how the dynamical organization of the emerging systems leads spontaneously to degree-frequency correlation at the node level, a structure typically associated to networks able to sustain explosive synchronization. 
Our results can widen our understanding of the shaping mechanisms behind the structural organization of some real-world systems such as brain networks where the emergence of explosive synchronization has been observed. The stability analysis of a simplified model reveals the microscopic mechanisms that are at the heart of the observed emerging structural and dynamical features, and their non-trivial correlations, observed in the networks described by our adaptive coevolving model.

\appendix*
\section{Stability analysis}

We compute the fixed points for Eq.~(\ref{fun:2nodes}), for which the following two cases need to be considered:
\begin{eqnarray}
\phi^* = \arcsin ( \Delta/\sigma_c ), \;\;\; && \alpha^* = 1; \label{fun:case1}\\
\phi^* = \arccos (2 p_c^2 -1), \;\;\; && \alpha^* = \frac{(\Delta/\sigma_c)}{2 p_c \sqrt{1-p_c^2}}. \label{fun:case2}
\end{eqnarray}

As it is well known for the Kuramoto model, for $\phi$ to be locked it is required that $\sigma_c \ge \Delta$. Otherwise, neither $\phi^*$ in Eq.~(\ref{fun:case1}) nor $\alpha^*$ in Eq.~(\ref{fun:case2}) are well defined. Notice that Eq.~(\ref{fun:case2}) implies that $\alpha^* \ge \Delta/\sigma_c$, since $0 \le p_c \le 1$, and the weight $\alpha$ is constrained to be in the unit interval. This is a consequence of Eq.~(\ref{fun:weight}) that keeps all weights in the interval unit once the set of initial weights are chosen in that range.

The eigenvalues of the Jacobian matrix for Eq.~(\ref{fun:case1}) are
\[ \lambda_1 =- \sigma_c \cos \phi^*, \hspace{5mm} \lambda_2 = \sqrt{\frac{1+ \cos \phi^*}{2}} -p_c. \]
Therefore, if $p_c < 1/\sqrt{2}$, we have a saddle point since $\lambda_1 <0$ and $\lambda_2 >0$. If $p_c > 1/\sqrt{2}$, there are two options depending on the critical coupling strength
\begin{equation}
\Sigma_1 := \frac{\Delta}{2 p_c \sqrt{1- p_c^2}}.
\label{fun:sigma1}
\end{equation}
If $\sigma_c > \Sigma_1$, we again find a saddle point with $\lambda_1 <0$ and $\lambda_2 >0$, but if $\Delta < \sigma_c < \Sigma_1$, the fixed point is stable.

The corresponding Jacobian matrix for the second fixed point, Eq.~(\ref{fun:case2}), yields two more complex eigenvalues:
\[ \lambda_{\pm} = \frac{A \pm \sqrt{A^2 +  B (\frac{1}{\sigma_c} - \frac{1}{\Sigma_1})}}{C}, \]
where $A := 2(1-2 p_c^2) \Delta$, $B := 16 p_c (1-p_c^2) \Delta^2$ and $C := 8 p_c \sqrt{1-p_c^2}$. Since $B$ and $C$ are always positive, we deduce that if $\sigma_c < \Sigma_1$, the fixed point is a saddle point with $\lambda_{-} <0$ and $\lambda_{+} >0$. If $\sigma_c > \Sigma_1$, the stability depends on a second critical value $\Sigma_2$ that determines when the radicand becomes zero. This value can be defined as
\begin{equation}
\frac{1}{\Sigma_2} := \frac{1}{\Sigma_1} - \frac{(1-2 p_c^2)^2}{4 p_c (1- p_c^2)}. \label{fun:sigma2}
\end{equation}
It can be verified easily that $\Sigma_2 > \Sigma_1$ for any value of $\Delta$ and $p_c$.

When $\Sigma_1 < \sigma_c < \Sigma_2$, we have either two positive eigenvalues, if $A>0$ (i.e., if $p_c < 1/\sqrt{2}$), or two negative eigenvalues, if $A<0$ (i.e., if $p_c > 1/\sqrt{2}$). On the other hand, if $\sigma_c > \Sigma_2$, the radicand is negative and, consequently, both eigenvalues are complex and the sign of the real part is given by the sign of $A$, positive for $p_c < 1/\sqrt{2}$ and negative for $p_c > 1/\sqrt{2}$.

\begin{acknowledgments}
Work partly supported by SEP-PRODEP (Mexico) through the Project UACOAH-PTC-322 DSA/103.5/15/7082. Work partly supported by the Spanish Ministry of Economy under projects FIS2013-41057-P and SAF2016-80240-P, and by GARECOM (Group of Research Excelence URJC-Banco de Santander). Authors acknowledge the computational resources and assistance provided by CRESCO, the supercomputing center of ENEA in Portici, Italy.
\end{acknowledgments}


%

\end{document}